\documentstyle[12pt]{article}
\def\be{\beta}
\def\ga{\gamma}
\def\de{\delta}
\def\ep{\epsilon}

\def\ph{\phi}

\def\De{\Delta}

\def\La{\Lambda}

\def\Ps{\Psi}

\def\prt{\partial}

\def\frac#1#2{{\textstyle{{#1}\over {#2}}}}

\def\lsim{\mathrel{\rlap{\lower4pt\hbox{\hskip1pt$\sim$}}
    \raise1pt\hbox{$<$}}}
\def\gsim{\mathrel{\rlap{\lower4pt\hbox{\hskip1pt$\sim$}}
    \raise1pt\hbox{$>$}}}
\def\sqr#1#2{{\vcenter{\vbox{\hrule height.#2pt
         \hbox{\vrule width.#2pt height#1pt \kern#1pt
         \vrule width.#2pt}
         \hrule height.#2pt}}}}

\newcommand{\beq}{\begin{equation}}
\newcommand{\eeq}{\end{equation}}
\newcommand{\bea}{\begin{eqnarray}}
\newcommand{\eea}{\end{eqnarray}}
\newcommand{\rf}[1]{(\ref{#1})}
 
\renewenvironment{thebibliography}[1]
 { \rm
  \begin{list}{\arabic{enumi}.}
    {\usecounter{enumi} \setlength{\parsep}{0pt}
     \setlength{\itemsep}{3pt} \settowidth{\labelwidth}{#1.}
     \sloppy
    }}{\end{list}}

\begin{document}
 
\begin{flushright}
{IUHET 384\\}
{January 1998\\}
\end{flushright}

\vglue 1cm
	    
\begin{center}
{{\bf TESTING CPT AND LORENTZ SYMMETRY WITH\\}
{\bf NEUTRAL-KAON OSCILLATIONS\\}
\vglue 1.0cm
{V. Alan Kosteleck\'y\\}
\bigskip
\small
{\it Physics Department, Indiana University\\}
{\it Bloomington, IN 47405, U.S.A.\\}
}
\vglue 0.8cm
 
\end{center}

\small
{\rightskip=3pc\leftskip=3pc\noindent
\baselineskip=12pt
In this talk, 
a brief summary is given of the possibility 
of using neutral-kaon oscillations
to bound CPT- and Lorentz-breaking parameters in
an extension of the minimal standard model.

}

\normalsize
\baselineskip=14pt
\vskip 1 cm

The product CPT of charge conjugation C, 
parity reflection P,
and time reversal T
is known to be a discrete symmetry 
in generic local relativistic quantum field theories of
point particles in flat spacetime,
including the standard model
\cite{cpt}.
Many experimental tests of CPT exist
\cite{pdg}.
At present,
the tightest published constraints on CPT violation
come from neutral-kaon interferometry.
For instance,
the experiment E773 at Fermilab
\cite{e773}
has published a bound
$r_K \equiv |m_K - m_{\overline{K}}|/m_K
< 1.3 \times 10^{-18}$
at the 90\% confidence level.

The attained precision of these tests
and the CPT invariance of the standard model 
make CPT breaking a candidate signal for new physics,
such as string theory
\cite{kp1}.
Indeed,
a theoretical framework for CPT and Lorentz violation does exist.
It is based on the notion that 
apparent low-energy CPT and Lorentz violation 
might emerge as a result of spontaneous symmetry breaking
in an underlying CPT- and Lorentz-invariant theory
\cite{ks,kp1}.
The potentially observable effects of CPT and Lorentz breaking
would then merely be a consequence of the vacuum structure,
which means many attractive properties  
of typical Lorentz-invariant theories 
would be preserved.

Effects of this type can be described at the level
of the minimal SU(3) $\times$ SU(2) $\times$ U(1) standard model
by including additional terms 
that could arise from spontaneous CPT and Lorentz breaking
but that preserve gauge invariances 
and power-counting renormalizability
\cite{ck}.
The existence of this explicit extension of the standard model 
means that possible experimental signals for 
apparent CPT and Lorentz breaking
can be investigated quantitatively. 
Limits on the breaking then take the form of 
bounds on coefficients of the new terms.

Using this approach,
various experimental consequences of 
possible CPT and Lorentz violation have been investigated.
The absence to date of observed violation 
suggests a high degree of suppression 
for possible effects at our present energies.
If the standard model is regarded as an effective theory
at the electroweak scale $m_{\rm ew}$
that arises from a fundamental theory
at the Planck scale $m_{\rm Pl}$,
then the natural suppression factor
for CPT and Lorentz violation would be
$r \sim m_{\rm ew}/m_{\rm Pl} \simeq 10^{-17}$.
The small size of this factor suggests  
only a few CPT- and Lorentz-violating effects
might be experimentally observable. 
However, 
some possible signals do exist.
They include,
for example,
ones in the 
$K$, $D$, $B_d$, and $B_s$
neutral-meson systems 
\cite{kp2,k,ck2,kvk,exptb},
in various QED contexts
\cite{ck,bkr},
and in baryogenesis
\cite{bckp}.
This talk focuses on tests with neutral mesons, 
in particular on the application of strangeness oscillations
in the neutral-kaon system as interferometric CPT tests.

Let the symbol $P$ denote any of the four neutral mesons
$K$, $D$, $B_d$, $B_s$.
In $P$-meson interferometry,
a phenomenological description of the time evolution 
is used to extract constraints on CPT violation. 
The Schr\"odinger wave functions 
of the strong-interaction particle eigenstate $P^0$ 
and its opposite-flavor antiparticle state $\overline{P^0}$
are taken to form a two-component object $\Ps$.
The time evolution is determined 
by the equation $i\prt_t \Ps = \La \Ps$,
where $\La$ is an effective hamiltonian.
Flavor oscillations between $P^0$ and $\overline{P^0}$
are governed by the off-diagonal components of $\La$.

A phenomenological parametrization of $\La_P$
permits two possible kinds of (indirect) CP violation.
The usual one is T violation with CPT invariance,
and the relevant parameter is denoted $\ep_P$.
In the kaon system,
a nonzero value of $\ep_K$ is observed
\cite{pdg},
and it can be related to CKM matrix elements
in the minimal standard model.
The other kind of CP violation is 
CPT violation with T invariance,
governed by a complex parameter 
denoted $\de_P$.
In the kaon system,
a constraint on $\de_K$
arises from a limit on $r_K$,
and $\de_K$ can be related to coefficients of the 
CPT-violating terms in the standard-model extension
\cite{ck}.

It can be shown that $\de_P$ is sensitive to only 
one type of term in the standard-model extension 
\cite{k}.
Denoting a quark field by $q$,
the relevant lagrangian term is
$- a^q_{\mu} \overline{q} \ga^\mu q$.
Here,
$a^q_{\mu}$ is constant in spacetime 
but depends on the quark flavor $q$.
No other lagrangian terms in the standard-model extension
affect $\de_P$.
Moreover,
neutral-meson interferometry 
is the only class of experiments known to be 
sensitive to the parameters $a_\mu$.

Since the $\La$ formalism is based on nonrelativistic
quantum mechanics,
the parameter $\de_P$ is defined in a frame comoving
with the $P$ meson.
Many experiments involve relativistic mesons,
and it can be shown that the presence of CPT and Lorentz breaking
implies that $\de_P$ is boost dependent.
Denoting the $P$-meson four-velocity by
$\be^\mu \equiv \ga(1,\vec\be)$,
a general expression for $\de_P$ is
\cite{k}
\beq
\de_P \approx i \sin\hat\ph \exp(i\hat\ph) 
\ga(\De a_0 - \vec \be \cdot \De \vec a) /\De m
\quad ,
\label{a}
\eeq
where 
$\De a_\mu \equiv a_\mu^{q_2} - a_\mu^{q_1}$
with $q_1$ and $q_2$ denoting the two flavors of valence
quark in the $P$ meson.
In Eq.\ \rf{a},
the mass and decay-rate differences
between $P_L$ and $P_S$ are,
respectively,
$\De m \equiv m_L- m_S$ and 
$\De \ga \equiv \ga_S- \ga_L$,
while $\hat\ph\equiv \tan^{-1}(2\De m/\De\ga)$.
A subscript $P$ is understood on all these quantities.
Equation \rf{a} holds at leading order
in all the Lorentz-breaking parameters
in the standard-model extension.

The above analysis has several implications for experiment.
The hermiticity of the standard-model extension 
means that $a_\mu$ is real.
This implies a proportionality
between the real and imaginary parts of $\de_P$,
which in principle provides an experimental signal
for spontaneous CPT and Lorentz breaking
\cite{kp2}.
Also,
the flavor dependence of the parameters $a_\mu^q$
means that the magnitudes of $\de_P$ for distinct $P$
could differ significantly,
so it is potentially important to test CPT symmetry 
in several neutral-meson systems.
Note that any direct CPT violation in $P$-meson decay amplitudes
would be suppressed by the factor $r\simeq 10^{-17}$
and hence would be too small to detect.

A significant feature of Eq.\ \rf{a}
is the appearance of a multiplicative factor of $\ga$.
This can enhance the CPT-violating effect for boosted mesons
\cite{k}.
For example,
under certain circumstances the CPT reach of an experiment
can be improved by increasing the boost,
as well as by increasing the statistics.
The momentum dependence also means that published limits on $\de_P$ 
from different experiments may represent inequivalent
CPT sensitivities.
Moreover,
typical experiments involve 
a spectrum of neutral-meson momenta.
The momentum dependence predicted by Eq.\ \rf{a}
therefore offers in principle a striking signal
for CPT and Lorentz breaking.

Equation \rf{a} also describes orientation-dependent effects,
in that $\de_P$ depends on the direction of 
propagation of the $P$ meson
according to the angle between $\vec \be$ and $\De \vec a$.
This has several implications
\cite{k}.
For example,
since bounds on $\de_P$ are typically extracted from data that are
taken over many days,
the Earth's rotation must be taken into account,
either by a suitable averaging or by time binning.
The type of experiment is also a relevant factor. 
Consider,
for example,
experiments with correlated $P$-$\overline P$ pairs 
generated via quarkonium decays.
For a symmetric factory with unboosted quarkonia
the momentum distribution is of lesser interest,
being essentially a line spectrum,
but a $4\pi$ coverage
would permit detailed studies of the angular dependence. 
In this case,
$\de_P$ and the associated CPT behavior 
of the two $P$ mesons in each correlated pair
differ because their velocities are opposite. 
In contrast,
in an asymmetric factory with boosted quarkonia
the $P$-meson momentum distribution is nontrivial
and can combine with the angular dependence 
to produce an interesting variety of CPT-violating effects.

\vglue 0.4cm

I thank Orfeu Bertolami, Robert Bluhm, Don Colladay, 
Rob Potting, Neil Russell, Stuart Samuel, 
and Rick Van Kooten for collaborations.
This work is supported in part
by the United States Department of Energy 
under grant number DE-FG02-91ER40661.

\vglue 0.4cm

\end{document}